\documentclass[reprint, superscriptaddress, twocolumn, amsmath, amssymb, aps, prb]{revtex4-2}
\usepackage{graphicx}
\usepackage{dcolumn}
\usepackage{bm}
\usepackage{placeins}
\usepackage{appendix}
\usepackage{upgreek}

\usepackage{verbatim}
\usepackage[usenames,dvipsnames]{xcolor}
 \usepackage{amsmath}
 \usepackage{amsfonts}
 \usepackage{amssymb}
 \usepackage[colorlinks=true,citecolor=blue,linkcolor=red]{hyperref}

 \usepackage{bbold}     
 \usepackage[makeroom]{cancel}  
 \usepackage{multirow}    
 \usepackage[normalem]{ulem}        
 \usepackage{array}
 \usepackage{pdfpages}
\makeatletter
 \AtBeginDocument{\let\LS@rot\@undefined}
 \makeatother

\begin{document}


\title{Hard superconducting gap in PbTe nanowires}

\author{Yichun Gao}
\email{equal contribution}
\affiliation{State Key Laboratory of Low Dimensional Quantum Physics, Department of Physics, Tsinghua University, Beijing 100084, China}

\author{Wenyu Song}
\email{equal contribution}
\affiliation{State Key Laboratory of Low Dimensional Quantum Physics, Department of Physics, Tsinghua University, Beijing 100084, China}

\author{Shuai Yang}
\email{equal contribution}
\affiliation{State Key Laboratory of Low Dimensional Quantum Physics, Department of Physics, Tsinghua University, Beijing 100084, China}

\author{Zehao Yu}
\affiliation{State Key Laboratory of Low Dimensional Quantum Physics, Department of Physics, Tsinghua University, Beijing 100084, China}

\author{Ruidong Li}
\affiliation{State Key Laboratory of Low Dimensional Quantum Physics, Department of Physics, Tsinghua University, Beijing 100084, China}

\author{Wentao Miao}
\affiliation{State Key Laboratory of Low Dimensional Quantum Physics, Department of Physics, Tsinghua University, Beijing 100084, China}

\author{Yuhao Wang}
\affiliation{State Key Laboratory of Low Dimensional Quantum Physics, Department of Physics, Tsinghua University, Beijing 100084, China}

\author{Fangting Chen}
\affiliation{State Key Laboratory of Low Dimensional Quantum Physics, Department of Physics, Tsinghua University, Beijing 100084, China}

\author{Zuhan Geng}
\affiliation{State Key Laboratory of Low Dimensional Quantum Physics, Department of Physics, Tsinghua University, Beijing 100084, China}

\author{Lining Yang}
\affiliation{State Key Laboratory of Low Dimensional Quantum Physics, Department of Physics, Tsinghua University, Beijing 100084, China}

\author{Zezhou Xia}
\affiliation{State Key Laboratory of Low Dimensional Quantum Physics, Department of Physics, Tsinghua University, Beijing 100084, China}

\author{Xiao Feng}
\affiliation{State Key Laboratory of Low Dimensional Quantum Physics, Department of Physics, Tsinghua University, Beijing 100084, China}
\affiliation{Beijing Academy of Quantum Information Sciences, Beijing 100193, China}
\affiliation{Frontier Science Center for Quantum Information, Beijing 100084, China}
\affiliation{Hefei National Laboratory, Hefei 230088, China}

\author{Yunyi Zang}
\affiliation{Beijing Academy of Quantum Information Sciences, Beijing 100193, China}
\affiliation{Hefei National Laboratory, Hefei 230088, China}

\author{Lin Li}
\affiliation{Beijing Academy of Quantum Information Sciences, Beijing 100193, China}

\author{Runan Shang}
\affiliation{Beijing Academy of Quantum Information Sciences, Beijing 100193, China}
\affiliation{Hefei National Laboratory, Hefei 230088, China}

\author{Qi-Kun Xue}
\affiliation{State Key Laboratory of Low Dimensional Quantum Physics, Department of Physics, Tsinghua University, Beijing 100084, China}
\affiliation{Beijing Academy of Quantum Information Sciences, Beijing 100193, China}
\affiliation{Frontier Science Center for Quantum Information, Beijing 100084, China}
\affiliation{Hefei National Laboratory, Hefei 230088, China}
\affiliation{Southern University of Science and Technology, Shenzhen 518055, China}

\author{Ke He}
\email{kehe@tsinghua.edu.cn}
\affiliation{State Key Laboratory of Low Dimensional Quantum Physics, Department of Physics, Tsinghua University, Beijing 100084, China}
\affiliation{Beijing Academy of Quantum Information Sciences, Beijing 100193, China}
\affiliation{Frontier Science Center for Quantum Information, Beijing 100084, China}
\affiliation{Hefei National Laboratory, Hefei 230088, China}

\author{Hao Zhang}
\email{hzquantum@mail.tsinghua.edu.cn}
\affiliation{State Key Laboratory of Low Dimensional Quantum Physics, Department of Physics, Tsinghua University, Beijing 100084, China}
\affiliation{Beijing Academy of Quantum Information Sciences, Beijing 100193, China}
\affiliation{Frontier Science Center for Quantum Information, Beijing 100084, China}


\begin{abstract}

Semiconductor nanowires coupled to a superconductor provide a powerful testbed for quantum device physics such as Majorana zero modes and gate-tunable hybrid qubits. The performance of these quantum devices heavily relies on the quality of the induced superconducting gap. A hard gap, evident as vanishing subgap conductance in tunneling spectroscopy, is both necessary and desired. Previously, a hard gap has been achieved and extensively studied in III-V semiconductor nanowires (InAs and InSb). In this study, we present the observation of a hard superconducting gap in PbTe nanowires coupled to a superconductor Pb. The gap size ($\Delta$) is $\sim$ 1 meV (maximally 1.3 meV in one device). Additionally, subgap Andreev bound states can also be created and controlled through gate tuning. Tuning a device into the open regime can reveal Andreev enhancement of the subgap conductance, suggesting a remarkable transparent superconductor-semiconductor interface, with a transparency of $\sim$ 0.96. These results pave the way for diverse superconducting quantum devices based on PbTe nanowires. 

\end{abstract}

\maketitle  

A superconductor can induce superconductivity in a semiconductor through proximity coupling. This coupling was spatially non-uniform in early devices due to disorder at the semiconductor-superconductor interface \cite{Takei2013}. While supercurrent can still be clearly resolved in Josephson devices \cite{Leo_Supercurrent}, the induced superconducting gap was soft with many quasi-particle states inside the gap. The soft gap can significantly degrade the quality of the corresponding device \cite{Mourik}, through mechanisms such as dissipation smearing or quasi-particle poisoning. Semiconductor nanowires provide a natural platform to study the induced superconducting gap, where an electrostatic gate can readily define a point contact for tunneling spectroscopy. The tunneling conductance reflects the density of states and can thus be associated with the quality of the induced superconducting gap. The issue of a soft gap in semiconductor nanowires can be addressed by suppressing interface disorder \cite{Gul2017, Zhang2017Ballistic}. Furthermore, epitaxial growth of superconductors can yield atomically sharp interfaces and hard gaps \cite{Chang2015, Krogstrup2015, Schapers_PRA, PanCPL}. These material improvements enable high quality quantum devices, e.g. on Majorana experiments \cite{Deng2016,Gul2018, Song2022, WangZhaoyu,Delft_Kitaev, MS_2023, NextSteps,Prada2020, cao2022recent} and hybrid qubits \cite{2015_PRL_gatemon, DiCarlo_gatemon,2019_PRX_Scalay,2021_Devoret_Science, Huo_gatemon, Delft_2023_qubit_NP}. So far, most of the hard gap studies are based on III-V semiconductors such as InAs or InSb. The Majorana theory \cite{Lutchyn2010, Oreg2010} requires a clean semiconductor nanowire while the current experiments are disorder-limited \cite{Patrick_Lee_disorder_2012, Loss2013ZBP, GoodBadUgly, DasSarma2021Disorder, Tudor2021Disorder}. 

PbTe nanowires have been recently proposed \cite{CaoZhanPbTe} to overcome the disorder issue, owing to their large dielectric constant ($\sim$ 1350). Meanwhile, these nanowires have been grown using selective area epitaxy \cite{Jiangyuying,Erik_PbTe_SAG} and subjected to quantum transport characterization \cite{PbTe_AB, Fabrizio_PbTe, Wenyu}. The superconducting proximity effect has also been demonstrated in PbTe-Pb Josephson junctions \cite{Zitong}. While a superconducting gap with a size of  $\Delta \sim$ 0.4 meV can be inferred, it is essential to note that tunneling spectroscopy in a Josephson device does not directly unveil the quasi-particle density of states (DOS) but rather a convolution of them. Moreover, the subgap conductance may have other contributions from the supercurrent and multiple Andreev reflections. Therefore, to directly probe the induced gap, N-NW-S devices are needed (where N stands for normal metal, NW for nanowire and S for superconductor). We have recently improved the material quality and here demonstrate the hard superconducting gap in PbTe nanowires in N-NW-S devices. We used Pb as the superconductor and measured a superconducting gap of $\Delta \sim$ 1 meV. This value is comparable to those observed in InAs-Pb devices \cite{InAs-Pb} and significantly larger than those reported in our early devices \cite{Zitong}. The transparency, estimated based on the Andreev enehancement in the open regime, can reach $\sim$ 0.96, also significantly higher than before ($<$ 0.8) \cite{Zitong}. Andreev bound states (ABSs) can be generated and controlled by gate voltages, providing an additional material platform for investigating Andreev physics.  These results are a pivotal advancement in utilizing PbTe nanowires for advanced quantum devices.

\begin{figure}[htb]
\includegraphics[width=\columnwidth]{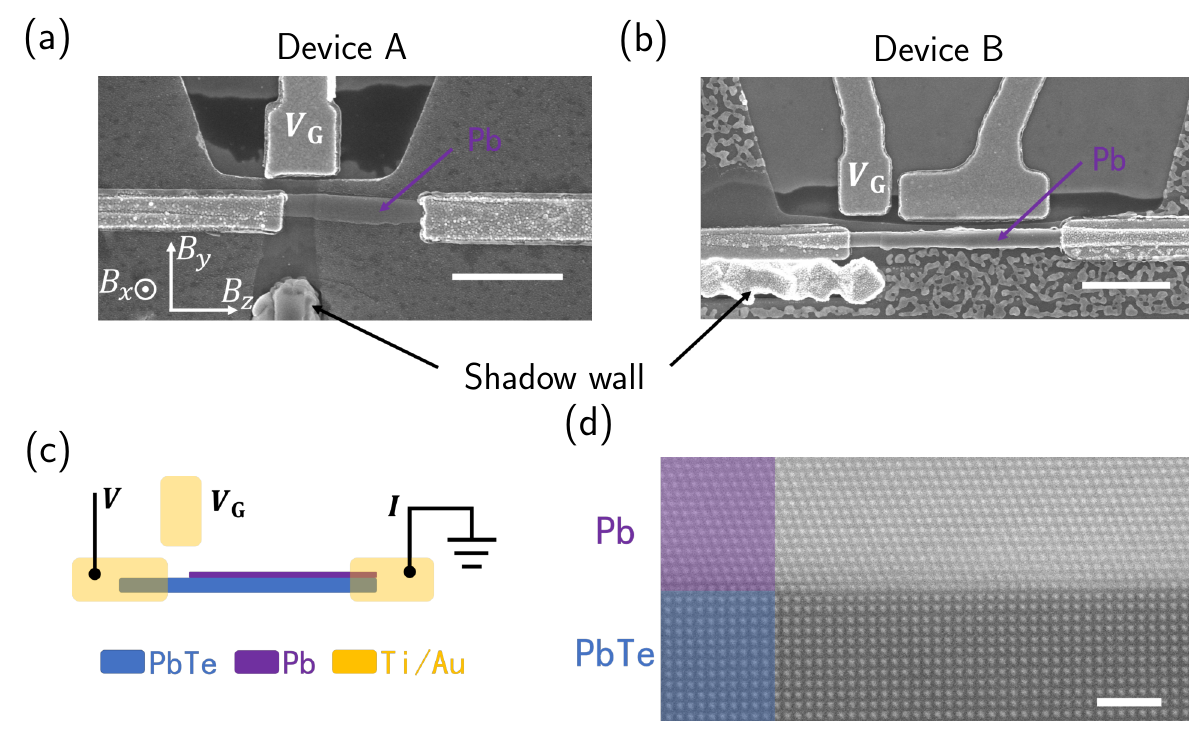}
\centering
\caption{Device basics. (a-b) SEMs of devices A and B, respectively. Scale bar, 1 $\upmu$m. The right gate in device B was kept grounded during the measurement. (c) Schematic of the device and measurement circuit: PbTe nanowire (blue), Pb superconductor (purple), contacts and gate (yellow). (d) STEM of a typical PbTe-Pb interface grown under identical conditions. Scale bar, 2 nm.}
\label{fig1}
\end{figure}

Figures 1(a) and 1(b) depict scanning electron micrographs (SEMs) of two devices (A and B). The PbTe nanowires were grown on a CdTe(001)/Pb$_{1-x}$Eu$_x$Te substrate, and subsequently covered epitaxially with a thin Pb film (thickness $\sim$ 10 nm) and a CdTe capping layer. The value of $x$ is estimated to be 0.01. The substrate was cooled using liquid nitrogen during the Pb growth process. The Pb film on the substrate of device B is discontinuous, likely due to the relatively high temperature during that particular growth. Nonetheless, the Pb films on the PbTe nanowires in both devices are uniform and continuous. The barrier region of the nanowire is defined by pre-patterned Hydrogen SilsesQuioxane (HSQ) shadow wall. The barrier transmission can be tuned by applying a voltage ($V_{\text{G}}$) to the side gate. The source and drain contacts, along with the side gates, were fabricated using evaporated Ti/Au (7/60 nm). Details of device fabrication can be found in the method section. Standard two-terminal measurement was performed in a dilution fridge at its base temperature (below 50 mK).  Figure 1(c) shows a simplified circuit and the schematic of the N-NW-S device. The scanning transmission electron microscopy (STEM) image in Fig. 1(d) displays a PbTe-Pb interface that was grown under identical conditions. An atomically sharp interface is crucial for the observation of a hard induced gap.

\begin{figure}[t]
\includegraphics[width=\columnwidth]{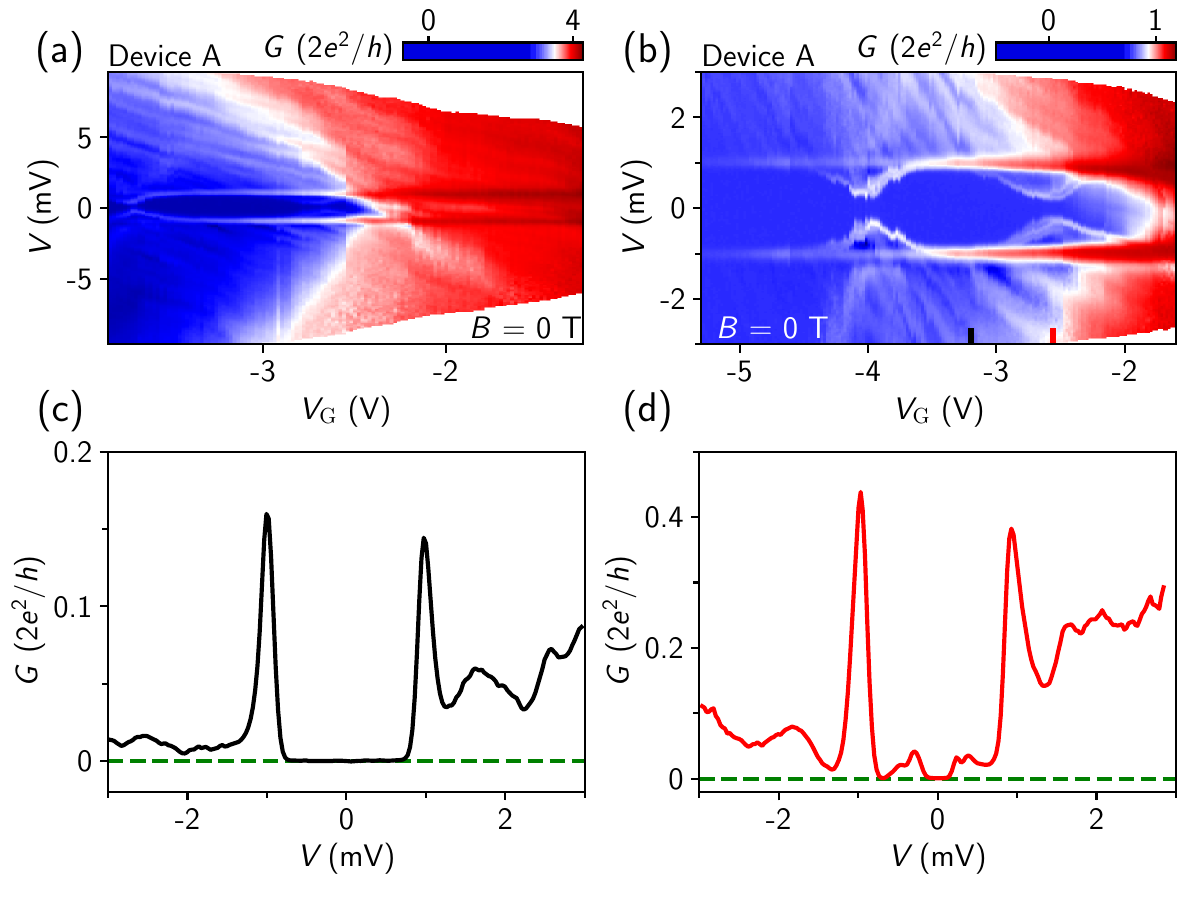}
\centering
\caption{Hard gap in device A. (a) $G$ vs $V$ and $V_{\text{G}}$ at zero field over a large $V$ range.  (b) A fine scan over a smaller $V$ range. (c) A line cut from (b) at $V_{\text{G}}$ = -3.2 V, see the black bar in (b). (d) A line cut of the subgap states at $V_{\text{G}}$ = -2.56 V, see the red bar in (b).  }
\label{fig2}
\end{figure}

\begin{figure*}[ht]
\includegraphics[width=0.65\textwidth]{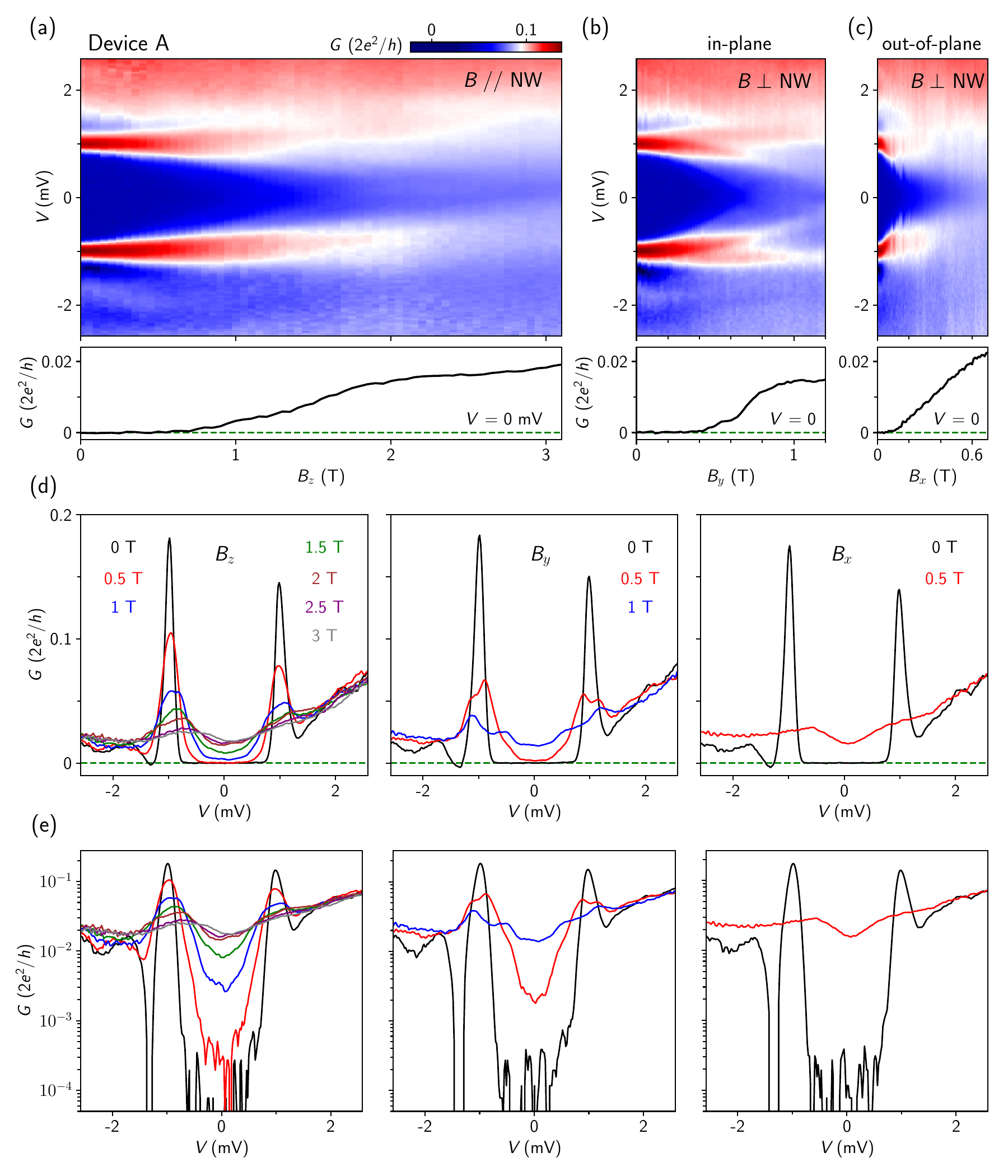}
\centering
\caption{$B$ dependence of the hard gap. (a) $G$ vs $V$ and $B_{z}$ (parallel to the nanowire). Lower panel, zero-bias line cut. (b-c) $B_{y}$ (in-plane) and $B_{x}$ (out-of-plane) scans of the gap.  Both are perpendicular to the nanowire. (d) Line cuts at different $B$'s from (a-c). (e) Log-scale plot of (d).  $V_{\text{G}}$ = -2.7 V for all panels.}
\label{fig3}
\end{figure*}

Figure 2(a) illustrates the conductance spectroscopy of device A over a large bias ($V$) range. The magnetic field ($B$) was kept at zero. The differential conductance, $G \equiv dI/dV$, was acquired using the standard lock-in technique. Note that the line resistance, mainly contributed by the fridge filters, has been subtracted from $G$. Additionally, $V$ has also been corrected by subtracting the bias drop over the line resistance. The two horizontal lines near $V = \pm 1$ mV are the coherence peaks that feature the edges of the superconducting gap. To better resolve the gap, Fig. 2(b) shows a fine scan over a smaller $V$ range. More negative $V_{\text{G}}$ depletes the barrier region and pinches off the device. Near the pinch-off regime, the tunneling conductance is proportional to the DOS of the proximitized PbTe. The line cut (Fig. 2(c)) shows sharp coherence peaks and vanishing sub-gap conductance, indicating a hard superconducting gap. The suppression of conductance (outside-gap vs sub-gap) can reach two orders of magnitude, see Fig. S1 for the log-scale plot in the Supplemental Material. The bias values of the coherence peaks suggest a gap size $\Delta \sim$ 1 meV, significantly larger than that of Al ($\sim$ 0.2$-$0.3 meV). A large superconducting gap is favorable for Majorana research as it sets an upper limit for the size of the topological gap. Larger gap can ``tolerate'' stronger disorder. Compared to the previous work on InAs-Pb nanowires \cite{InAs-Pb}, the gap size is similar but the coherence peaks here seem more sharply defined. Sharper coherence peaks suggest less dissipation smearing arising from undesirable quasi-particles \cite{Dynes}. 

Besides the hard gap, another notable feature is the discrete sub-gap levels at particular gate voltages. These gate-tunable levels are ABSs, revealed as sub-gap conductance peaks (see Fig. 2(d) for a line cut). ABSs are widely present in hybrid devices \cite{Pillet2010, Mason2011, Silvano2014, ZhangShan, WangZhichuan} and can mimic Majorana signatures, causing the decade-long debate \cite{Prada2012, BrouwerSmooth, Liu2017, Loss2018ABS,TudorQuasi, WimmerQuasi, CaoZhanPRL}. ABSs usually originate in quantum dots coupled to superconductors. Figures 2(a-b) show no obvious Coulomb blockades except for single resonant levels near $V_{\text{G}}$ = -4 V and $V_{\text{G}}$ = -2.5 V.  The white boundaries demarcating the red and blue regions in the color map suggest diamond like shapes, see the dashed lines in Fig. S1(a). These diamonds are unlikely to be Coulomb blockades, considering that the charging energy in PbTe should be two orders of magnitude smaller \cite{Fabrizio_PbTe}. The diamond slopes likely correspond to the subband bottoms of a quantum point contact \cite{Wenyu}. In Fig. S1 we show possible indication of quasi-ballistic transport after moderate smoothing on Fig. 2(a). In the absence of quantum dots, the observed ABSs are likely formed due to the smooth potential variation \cite{Prada2012, BrouwerSmooth}.

The outside-gap conductance exhibits Fabry-Perot-like oscillations over $V_{\text{G}}$, possibly caused by the scattering between the barrier region and the drain contact. These oscillations can affect the estimation of the normal state conductance ($G_{\text{N}}$). In Fig. S1, we present the extracted $G_{\text{N}}$ plotted against $G_{\text{S}}$, i.e. $G(V=0)$. We also compare it with the BTK-Beenakker formula \cite{BTK, Beenakker} for illustrative purposes. Deviation between the formula and experiment exist and may arise from various factors. Firstly, the aforementioned oscillations make the estimation of $G_{\text{N}}$ inaccurate. Secondly, the gap size of 1 meV is a significant fraction of the subband spacing in PbTe \cite{Wenyu}. Unlike the scenario of Al with a much smaller gap, the outside-gap (large-bias) conductance in PbTe-Pb may not accurately reflect the barrier transmission (normal state conductance) at zero bias. Lastly, a crucial factor is that the BTK-Beenakker formula assumes a single channel in the barrier, which may not always hold. Moreover, even for the single channel case, the formula only holds for high chemical potentials \cite{DasSarma2017QZBP}, thus has limited implications. 

\begin{figure}[ht]
\includegraphics[width=\columnwidth]{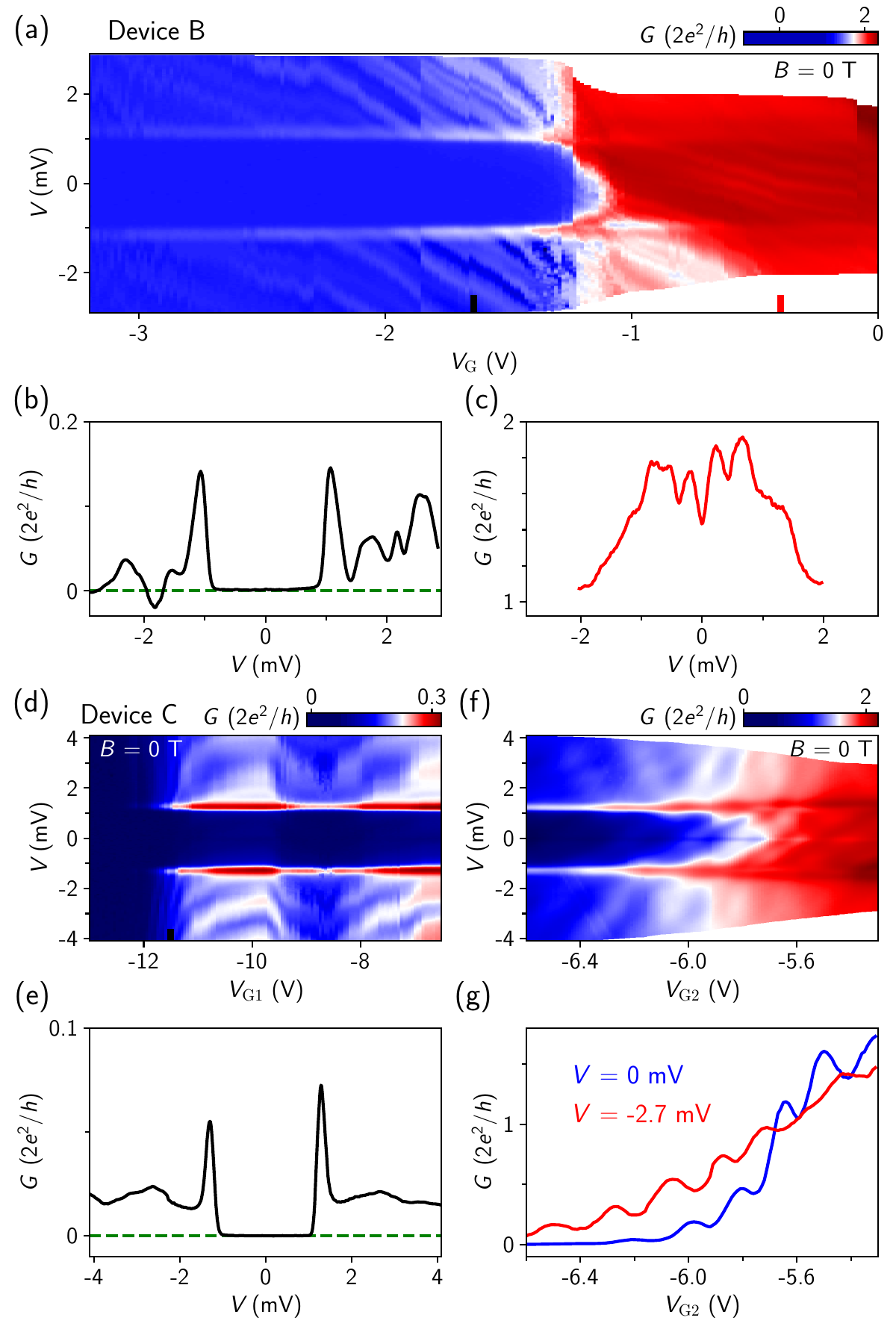}
\centering
\caption{Hard gap in devices B and C. (a) $G$ vs $V$ and $V_{\text{G}}$ in device B at zero field. (b) A line cut from (a) at $V_{\text{G}}$ = -1.64 V (the tunneling regime). (c) A line cut from (a) at $V_{\text{G}}$ = -0.395 V (the open regime). (d) Hard gap spectroscopy of device C. $V_{\text{G2}}$ = 0 V. $V_{\text{G1}}$ and $V_{\text{G2}}$ are the voltages on the two tunnel gates. (e)  A line cut from (d) at $V_{\text{G1}}$ = -11.5 V. (f) $G$ vs $V$ and $V_{\text{G2}}$ in the more opened regime. $V_{\text{G1}}$ = -4.8 V. (g) Horizontal line cuts from (f) at $V$ = 0 mV (blue) and -2.7 mV (red), respectively. }
\label{fig4}
\end{figure}
 
We next study the $B$ dependence of the hard gap (for the $B$ dependence of ABSs, see Fig. S2 in the Supplemental Material). Figures 3(a-c) show the $B$ scans along three coordinate axes, see Fig. 1(a) for the labeling. $B_z$ is roughly parallel to the nanowire (with an accuracy within 10$^\circ$). The gap becomes soft for $B_{z}$ approaching 1 T, see the lower panel for the zero-bias line cut. The gap closes as $B_z$ approaches 2 T. The other two directions, $B_{y}$ (in-plane) and $B_{x}$ (out-of-plane), are both perpendicular to the nanowire. In the $B_{y}$ scan, the gap closes at a lower field ($\sim$ 1 T) than the $B_z$ case, while in the $B_{x}$ scan, an even lower field ($\sim$ 0.5 T) is sufficient to close the gap. We attribute the gap closing in the $B_x$ scan to the orbital effect, given that $B_x$ is perpendicular to the Pb film. As for $B_{z}$ and $B_{y}$, the orbital effect is smaller, as both directions are in-plane (parallel to the Pb film). We ascribe the anisotropy between $B_z$ and $B_y$ to spin-orbit interaction. $B_{z}$ aligns with the nanowire, therefore is perpendicular to the spin-orbit direction ($B_{\text{SO}}$). Spin-orbit interaction protects the gap from closing  when $B$ is perpendicular to $B_{\text{SO}}$, thus a higher critical field is observed for $B_z$. $B_{y}$ is parallel to $B_{\text{SO}}$. Hence the gap closes without the protection and a lower critical field is observed. This spin-orbit-induced anisotropy has been previously demonstrated in InSb nanowires coupled to a superconductor NbTiN \cite{Jouri2019}.

Figures 3(d) and 3(e) illustrate line cuts at different fields in linear and logarithmic scales for these three directions to further illustrate this anisotropy. At zero field, the gap shows a conductance suppression over two orders of magnitude. At 0.5 T (the red curves), the gap still remains hard in the $B_{z}$ scan while the sub-gap conductance has already been increased by an order of magnitude in the $B_{y}$ scan and by two orders of magnitude in the $B_{x}$ scan. In Fig. S3 in the Supplemental Material, we present a similar $B$ anisotropy observed in another device. Note that in the InAs-Pb work \cite{InAs-Pb}, the hard gap can persist to a considerable higher $B$ ($\sim$ 8.5 T). Further increasing the critical field for PbTe nanowires should be achievable and can be a topic for our forthcoming optimization efforts. 

For reproducibility purpose, we present two additional devices (B and C) in Fig. 4 demonstrating hard gaps. One more device (device D) can also be found in Fig. S3. Figure 4(a) shows the conductance map of device B. In the tunneling regime, a gap of $\Delta \sim$ 1.05 meV is observed, see Fig. 4(b) for a line cut. The negative differential conductance outside the gap (near the coherence peak) is typical in nanowire hybrids with discrete energy levels \cite{Gul2017}. In the open regime where the outside-gap conductance is close to $2e^2/h$, the subgap conductance is enhanced relative to the outside-gap conductance, as shown Fig. 4(c). This enhancement is due to Andreev reflection. The enhancement factor is $\sim$ 1.7. Using the BTK-Beenakker formula $G_{\text{S}}=1.7\times2e^2/h=\frac{2T^2}{(2-T)^2}\times 2e^2/h$ \cite{BTK, Beenakker}, we can estimate the interface transparency $T \sim$ 0.96.

Device C is an island device \cite{Albrecht} but operated as a tunneling spectroscopy device by setting one barrier region to be open (conductance much larger than $2e^2/h$). A hard gap with sharp coherence peaks can be observed in Figs. 4(d-e). The gap size is $\sim$ 1.3 meV. For all devices (A, B, C and D), Fabry-Perot-like oscillations outside the gap are present. The oscillations for subgap conductance becomes noticeable in a more opened regime as shown in Figs. 4(f-g) for device C. The quasi-periodic oscillations suggest two well-defined scattering points, likely the two barrier regions of the island device. Note that even if one barrier is tuned to be considerably open, it can still act as a scattering point in Fabry-Perot oscillations.   

In summary, we have observed a hard superconducting gap with sharp coherence peaks in PbTe nanowires coupled to a superconductor Pb. The gap size is $\sim$ 1 meV, several times larger than that of Al. Our result may enable the search for high quality Majorana signatures in PbTe nanowires, as well as other PbTe-based hybrid superconducting quantum devices.

\textbf{Acknowledgment} This work is supported by Tsinghua University Initiative Scientific Research Program, National Natural Science Foundation of China (92065206) and the Innovation Program for Quantum Science and Technology (2021ZD0302400). Raw data and processing codes within this paper are available at  https://doi.org/10.5281/zenodo.8313595.

\bibliography{mybibfile}

\newpage

\onecolumngrid

\newpage


\section*{Supplementary material}

# Supplemental Material for "Hard superconducting gap in PbTe nanowires"

Yichun Gao,[1,][*] Wenyu Song,[1,][*] Shuai Yang,[1,][*] Zehao Yu,[1] Ruidong Li,[1] Wentao Miao,[1] Yuhao Wang,[1] Fangting Chen,[1] Zuhan Geng,[1] Lining Yang,[1] Zezhou Xia,[1] Xiao Feng,[1, 2, 3, 4] Yunyi Zang,[2, 4] Lin Li,[2] Runan Shang,[2, 4] Qi-Kun Xue,[1, 2, 3, 4, 5] Ke He,[1, 2, 3, 4,][†] and Hao Zhang[1, 2, 3,][‡]

[1]*State Key Laboratory of Low Dimensional Quantum Physics,*
*Department of Physics, Tsinghua University, Beijing 100084, China*
[2]*Beijing Academy of Quantum Information Sciences, Beijing 100193, China*
[3]*Frontier Science Center for Quantum Information, Beijing 100084, China*
[4]*Hefei National Laboratory, Hefei 230088, China*
[5]*Southern University of Science and Technology, Shenzhen 518055, China*

**Additional information about the device fabrication and measurement**

Growth of PbTe-Pb nanowires has been reported in our previous work [1]. The process here is similar but with modifications, which will be discussed elsewhere. To fabricate a device, most of the Pb film on the substrate needs to be etched away to prevent a short circuit. Firstly, the core regions of the SAG nanowires were manually coated by a resist (PMMA C5). The Pb film in the non-coated regions was etched via Ar ion milling for 162 s. For every 27 s of ion milling, we stopped for 2 min to prevent heating. The resist was then dissolved in Acetone. Next, the device chip was spin coated by PMMA C5 twice (bilayer) and vacuum pumped overnight. Note that to prevent the degradation of the Pb film, no hot plate baking was used. An electron beam lithography (EBL) was then performed to define and etch the regions of Pb film near the device. The Ti/Au contacts and gates were evaporated in another EBL process. Before the evaporation, an Ar plasma etching of 130 s was performed in situ for ohmic contacts. The measurement was standard two terminal, which has been detailed in our previous work [2, 3]. We have assumed a contact resistance of 300 Ω for device A and 0 Ω for other devices when calculating the series resistance of the circuit.

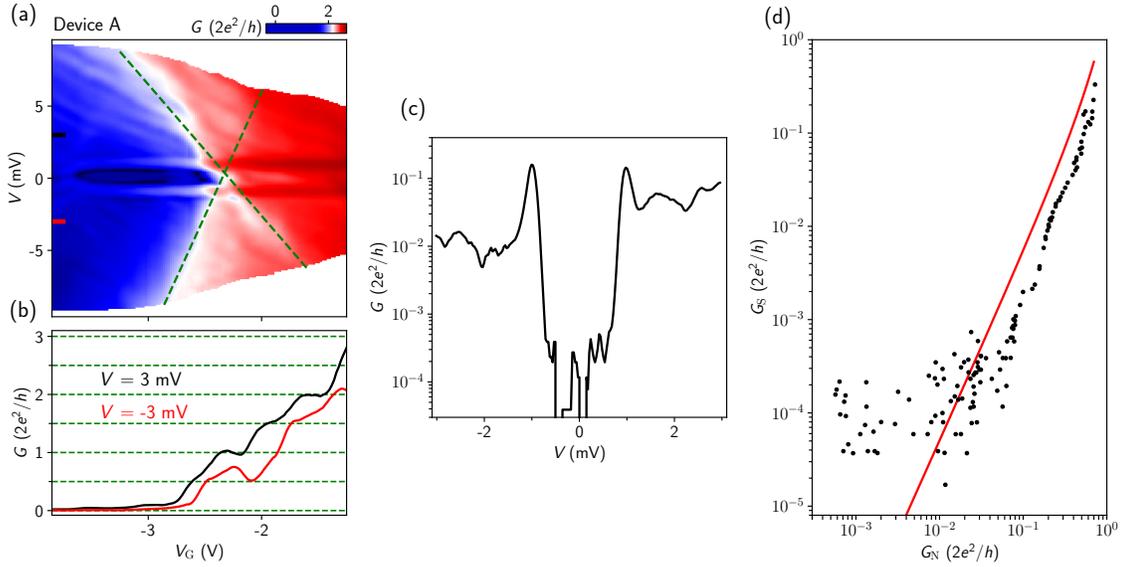

FIG. S1. (a) Replot of Fig. 2(a) with a moderate smoothing. The dashed lines are guide for eyes, indicating subband edges. (b) A horizontal line cut at $V = 3$ mV, resolving plateau-like features. Note that these "plateaus" can be affected by the Fabry-Perot oscillations, e.g. at other bias values (see the red curve at $V = -3$ mV). Device A is thus not fully ballistic, but close to the quasi-ballistic regime. (c) Replot of Fig. 2(c) in the logarithmic scale. The subgap vs outside gap suppression reaches two orders of magnitude. (d) $G_N$ vs $G_S$, extracted from Fig. 2(b). $G_S$ is the zero-bias conductance while $G_N$ is the average conductance over a bias window ($|V|$ between 1.3 mV and 1.4 mV). The red curve is calculated based on the BTK-Beenakker formula: $G_S = 2G_N^2/(2 - G_N)^2$ ($G_S$ and $G_N$ are in units of $2e^2/h$). The deviation between the experiment and the formula has been discussed in the maintext.

[*] equal contribution
[†] kehe@tsinghua.edu.cn
[‡] hzquantum@mail.tsinghua.edu.cn

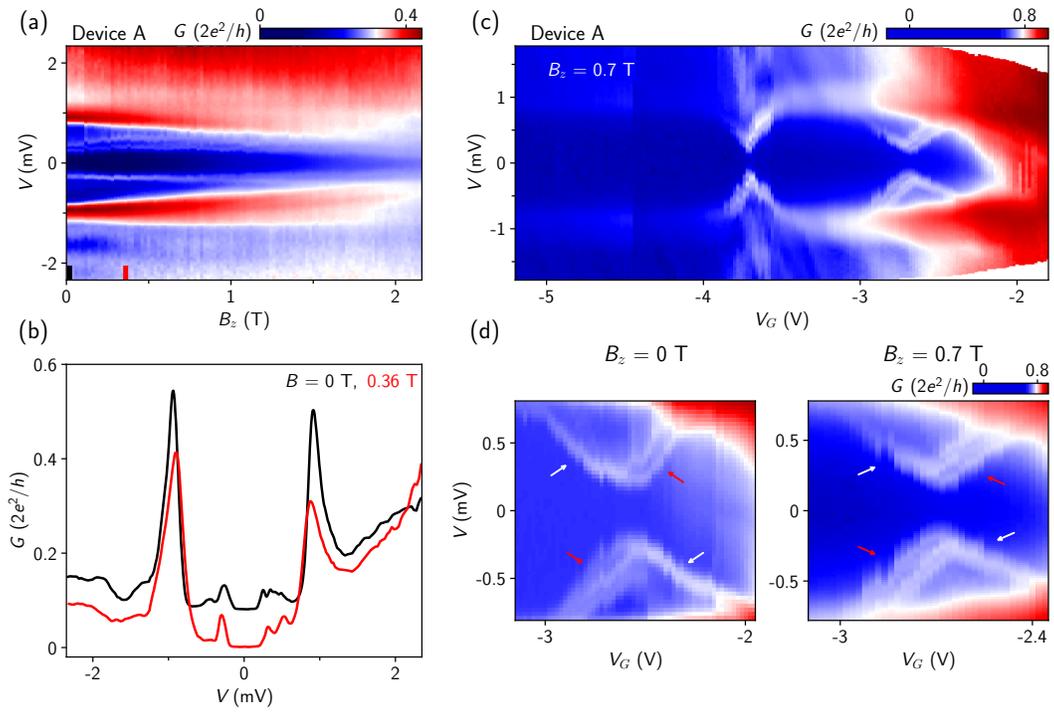

FIG. S2. (a) $B_z$ dependence of the ABS peaks. $V_G = -2.59$ V. The peaks initially move toward higher energies, suggesting a doublet ground state origin [4]. (b) Two line cuts from (a) at 0 T (with a vertical offset of $0.08 \times 2e^2/h$) and 0.36 T. (c) $V_G$ scan of the ABSs at $B_z = 0.7$ T. (d) Zoom-in of the ABSs at 0 T (from Fig. 2(b)) and 0.7 T (from panel (c)), respectively. The white arrows mark the ABS regions with a single peak while the red arrows for double or even triple peaks. This behavior does not fit into the simple singlet-doublet ABS picture [4] and suggests a more sophisticated ABS mechanism.

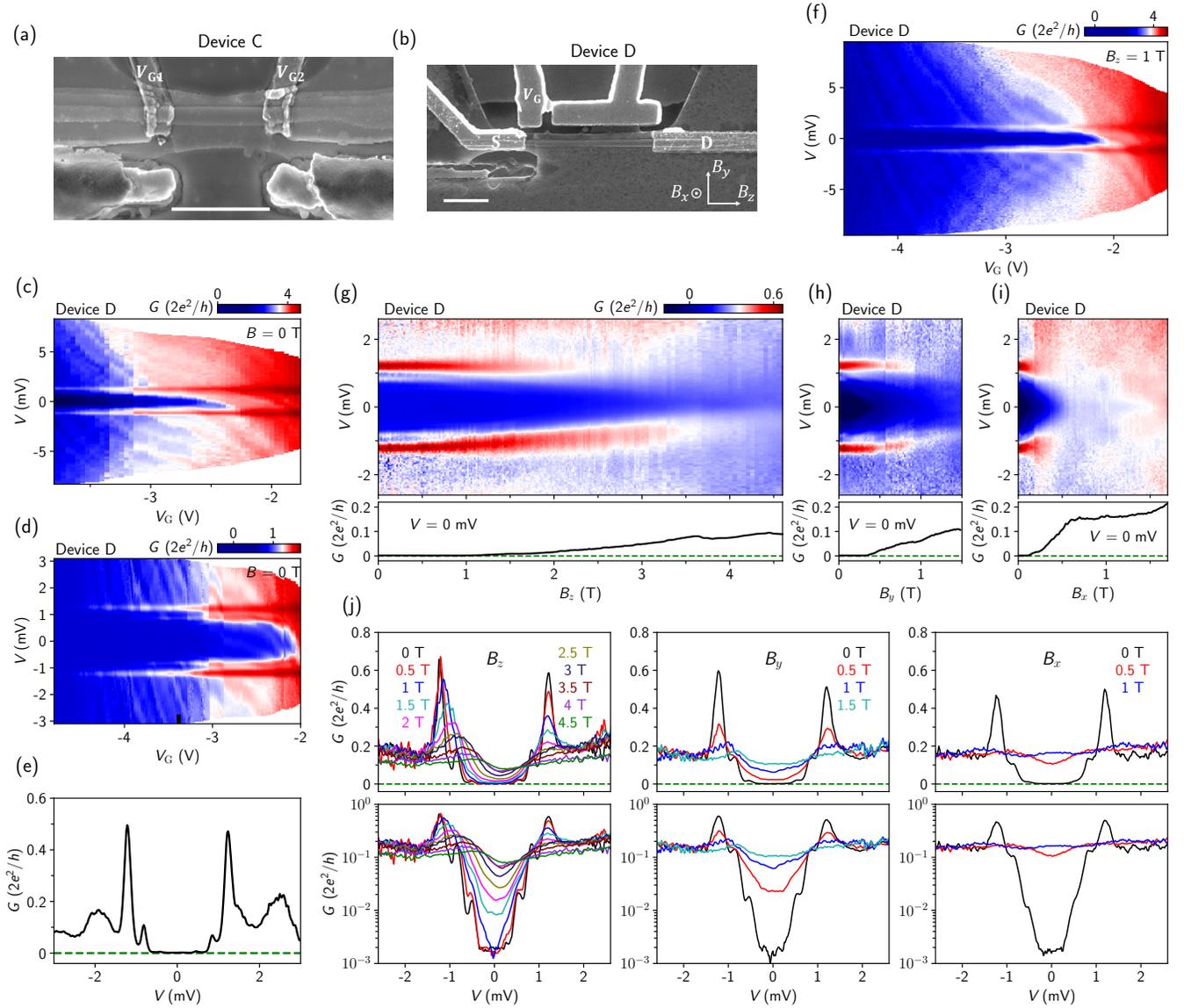

FIG. S3. (a-b) SEMs of devices C and D. Scale bar, 1 μm. The two tunnel gate voltages ($V_{G1}$ and $V_{G2}$) are labeled for device C. A dielectric layer of $HfO_2$ was sputtered before the top gate deposition for device C. The two gates of device D were shorted during the measurement. (c-d) Conductance 2D map of device D over large (c) and small (d) bias ranges, respectively. (e) A line cut from (d), see the black bar, exhibiting a hard superconducting gap. (f) Conductance spectroscopy at $B_z = 1$ T in device D. (g-i) $B$ dependence of the gap along three different directions for device D. Lower panels, zero-bias line cuts. (j) Line cuts from (g-i) in linear and logarithmic scales. The subgap vs outside gate conductance suppression can reach two orders of magnitude. The hard gap can persist to $\sim 1$ T for $B_z$. The anisotropy of gap closing along different $B$ directions in device D is similar to that in device A (Fig. 3).



\end{document}